\documentclass[letter]{aa} 

\usepackage{graphicx}
\usepackage{txfonts}
\usepackage{sidecap}

\newcommand{\gsim}{\hbox{\rlap{\lower.55ex\hbox{$\sim$}} \kern-.3em
\raise.4ex \hbox{$>$}}}
\newcommand{\lsim}{\hbox{\rlap{\lower.55ex\hbox{$\sim$}} \kern-.3em
\raise.4ex \hbox{$<$}}}

\newcommand{\hb}{H$\beta$}

\newcommand{\ha}{H$\alpha$}

\newcommand{\hei}{He\,\textsc{i}}

\newcommand{\nii}{\textsc{[N\,ii]}}

\newcommand{\oiii}{\textsc{[O\,iii]}}

\begin{document} 

\title{Towards DIB mapping in galaxies beyond 100~Mpc}
\subtitle{A radial profile of the $\lambda$5780.5 diffuse interstellar band in AM\,1353-272 B\thanks{Based on observations made with ESO telescopes at the La Silla Paranal Observatory under program ID 60.A-9100(B)}
}
   \author{A. Monreal-Ibero
          \inst{1}
          \and
          P. M. Weilbacher
          \inst{2}
          \and
          M. Wendt
          \inst{3,2}
         \and
         F. Selman
        \inst{4}
        \and
        R. Lallement
          \inst{1}
          \and
          \\
          J. Brinchmann
          \inst{5}
          \and
          S. Kamann
          \inst{6}
          \and
          C. Sandin
             \inst{2}       
          }
   \institute{GEPI, Observatoire de Paris, CNRS, Universit\'e Paris-Diderot, Place Jules Janssen, 92190 Meudon, France \\
              \email{ana.monreal-ibero@obspm.fr}
              \and
             Leibniz-Institut f\"ur Astrophysik Potsdam (AIP), An der Sternwarte 16, 14482 Potsdam, Germany 
             \and
             Institut f\"ur Physik und Astronomie, Universit\"at Potsdam, D-14476 Golm, Germany
             \and
             European Southern Observatory, 3107 Alonso de C\'ordova, Santiago, Chile 
             \and
             Leiden Observatory, Leiden University, P.O. Box 9513, NL-2300 RA Leiden, The Netherlands
             \and
             Institut f\"ur Astrophysik, Universit\"at G\"ottingen, Friedrich-Hund-Platz 1, 37077 G\"ottingen, Germany 
            }

   \date{Received  10 February 2014 / Accepted 20 February 2014}

 
  \abstract
   {Diffuse interstellar bands (DIBs) are non-stellar weak absorption features of unknown origin found in the spectra of stars viewed through one or several clouds of the interstellar medium (ISM). Research of DIBs outside the Milky Way is currently very limited. In particular, spatially resolved investigations of DIBs outside of the Local Group are, to our knowledge,  inexistent.}
   {In this contribution, we explore the capability of the high-sensitivity Integral Field Spectrograph, MUSE, as a tool for mapping diffuse interstellar bands at distances larger than 100~Mpc.}
   {We used MUSE commissioning data for  \object{AM\,1353-272}~ B, the member with the highest extinction of the Dentist's Chair, an interacting system of two spiral galaxies. High signal-to-noise spectra were created by co-adding the signal of many spatial elements distributed in a geometry of concentric elliptical half-rings.}
   {We derived decreasing radial profiles for the equivalent width of the $\lambda$5780.5 DIB both in the receding and approaching side of the companion galaxy up to distances of $\sim$4.6~kpc from the centre of the galaxy.
    The interstellar extinction as derived from the \ha/\hb\ line ratio displays a similar trend, with decreasing values towards the external parts. 
   This translates into an intrinsic correlation between the strength of the DIB and the extinction within \object{AM\,1353-272}~B, consistent with the currently existing global trend between these quantities when using measurements for Galactic and extragalactic sightlines.}
   {It seems feasible to map the DIB strength in the Local Universe,
which has up to now only been performed for the Milky Way. This offers a new approach to studying the relationship between DIBs and other characteristics and species of the ISM in addition to using galaxies in the Local Group or sightlines towards very bright targets outside the Local Group.
 }


   \maketitle
%

\section{Introduction}

Diffuse interstellar bands (DIBs) are non-stellar weak
absorption features found in the spectra of stars viewed trough one or several clouds of
the interstellar medium  \citep[ISM,][]{Herbig95}.
They were identified for the first time by \citet{Heger22}, but their interstellar origin was established in the 1930s \citep{Merrill34}.
Almost one century after their discovery, the nature of their
carriers (i.e. the agent that causes these features) remains a mystery \citep{Fulara00}. However, several of the DIBs are relatively strong and present good correlations with the amount of neutral hydrogen, the extinction, and the interstellar Na I D and Ca H and K
lines  along a given line of sight \citep[e.g.][]{Herbig93,Friedman11}. Thus, irrespective of the actual nature of carriers, DIBs can be used as tools to infer
the properties of the ISM structure, as shown by several recent works for the  Milky Way  \citep[e.g.][]{Puspitarini15,Zasowski15,Kos14,Cordiner13}.

Equivalent examples for other galaxies are difficult to find.
There is a certain level of spatial resolution in the investigations of DIBs in galaxies within the Local Group \citep{Ehrenfreund02,Cordiner11,vanLoon13}, since several sightlines in a given galaxy are sampled. Works targeting DIBs outside the Local Group are still rare and address individual sightlines for a given target \cite[e.g.][]{Heckman00,Phillips13}.
Currently, no mapping of a DIB has been performed for galaxies outside of the Local Group (beyond $\sim1$ Mpc).

The advent of highly efficient integral field units renders this feasible for the first time.
They simultaneously record the spatially resolved spectral information in an extended continuous field. This can be tessellated \emph{a posteriori}, preserving the relevant spatial information as best possible, but at the same time optimizing the quality of the signal by co-adding the spectra belonging to a given portion of the mapped area - the tile - \citep[e.g.][]{MonrealIbero05,Sandin08}. Here, we  use this principle to explore the capabilities of MUSE to obtain spatially resolved information of extragalactic DIBs.

We carried out our experiment in the interacting system \object{AM~1353-272}. The system consists of two components. The main galaxy (A) presents two prominent $\sim$40~kpc long tidal tails and was thoroughly studied by \citet{Weilbacher00,Weilbacher02,Weilbacher03}.
The companion (B) is a low-luminosity ($M_B=-18.1$\,mag) disk-like galaxy of disturbed morphology  undergoing a strong starburst and with high extinction. The relative positions of A and B are shown in Fig. 1 of \citet[][]{Weilbacher02}.
Here, we used \object{AM~1353-272}\,B as a test bench to explore the possibility of DIB detection and mapping.
We assume a distance of $D=159$~Mpc for AM~1353-272 using $H_0=75$~km~s$^{-1}$, as in \citet{Weilbacher02}. This implies a linear scale of a 0.771 kpc~arcsec$^{-1}$. 

\section{Observations and data reduction}

The interacting system \object{AM\,1353-272} was observed as part of commissioning run 2a \citep{Bacon14} of the MUSE instrument at the VLT.
Six 900\,s exposures were taken in two positions between 2014-04-29T04:22:00 and 06:00 under conditions of a few thin clouds and an average effective airmass of 1.04.
The seeing as given by the Gaussian FWHM of the stars in the white-light image of the final cube is 0\farcs8.
Each position was observed at position angles (P.A.) of 0$^\circ$, 90$^\circ$, and 180$^\circ$.
The instrument was set to wide-field mode ($\sim1\arcmin\times1\arcmin$ field of view) with a nominal wavelength range ($\sim 4800 \dots 9300 \AA$).
 The instrument resolution at $7000\AA$ is $R\sim2800$.
The standard star GJ\,754.1A was observed on
2014-04-28T09:57:00, and the reduction made use of twilight skyflats taken on 2014-04-27.

Standard reduction steps were followed with MUSE pipeline v1.0 \citep{Weilbacher14},
with bias subtraction, flat-fielding, and spectral tracing using the lamp-flat exposures, wavelength calibration, all
using standard daytime calibration exposures made the morning after the observations. The standard geometry table and astrometric solution for commissioning run 2a were
used to then transform the science data into pixel tables. This last step
included correction for atmospheric refraction, flux calibration, and sky
subtraction. 
The sky spectrum was measured in the blank areas of each exposure, set to the darkest 40\% of the field of view,
and data were corrected to barycentric velocities, resulting in corrections of -0.6 to -0.7 km s$^{-1}$ for the time of observations.
  The final, combined datacube was then reconstructed from all six
  exposures, sampling the data at $0\farcs2\times0\farcs2\times0.8\,\AA$,
  rejecting cosmic rays in the process.
We used a subset of 111$\times$91 spatial elements for the experiment presented here.

\section{Extracting information}

\subsection{Tile definition}

\begin{SCfigure}
   \includegraphics[width=0.30\textwidth, bb=30 90 475 453, clip=]{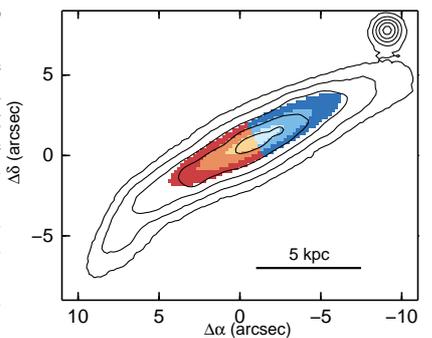} 
    \caption{Apertures used  to extract the high signal-to-noise spectra (i.e. the tiles). Each tile has been coloured differently using a palette that follows the blue-to-red velocity distribution within the galaxy.
    As a reference, the reconstructed white-light image is overplotted as contours in logarithmic stretching with 0.25~dex steps.
    }
     \label{figaper}
   \end{SCfigure}

The galaxy has  receding and approaching velocities at the southeast and northwest side, respectively (Weilbacher et al. in prep.).
Since the velocity gradient is strong ($\Delta v\gsim300$~km~s$^{-1}$), we treated each side of the galaxy independently, this way avoiding a large artificial broadening of the spectral features. To do
this, the tiles were defined as elliptical half rings following the morphology of the galaxy, with a fixed P.A.$=153^\circ$, measured counterclockwise, relative to the north, and an ellipticity of $b/a=0.2$.
The ring sizes represent a compromise between retaining as much   spatial resolution as possible and having a reliable DIB detection. The final tiles (depicted in Fig. \ref{figaper}) have between 29 and 159 spatial elements  each (projected areas between 0.7 and 3.8 kpc$^2$).

%
%

\subsection{Feature fitting}
\label{secfitting}

The spectra belonging to each tile were co-added, extracted, and corrected from a Galactic extinction of $A_v = 0.165$ \citep{Schlafly11} using the IRAF task \texttt{deredden} and assuming $R_V=A(V)/E(B-V)=3.1$ \citep{Rieke85}.
In each spectrum, we fitted different spectral features needed for our analysis using the IDL-based routine \texttt{mpfit} \citep{Markwardt09}.
The results discussed here are based on the equivalent width of the DIB at $\lambda$5780.5 (EW($\lambda$5780.5)), and fluxes for \ha, \hb, \oiii$\lambda$5007, and \nii$\lambda$6584.
We refer to \citet{MonrealIbero10a} for a description of the information recovery for  the emission lines
since it is quite standard and not particularly challenging for the experiment described here.

The most critical feature is the DIB at $\lambda$5780.5.
This was fitted together with the sodium doublet at $\lambda\lambda$5889.9,5895.9 and the  \hei$\lambda$5876 emission line using Gaussian functions, fixing their relative central positions according to a redshift of $z=0.039467$ ($v=11\,840$~km~s$^{-1}$ ), as determined from the average of the \ha\ centroid in the two inner half rings and in accord with the $11\,790\pm50$~km~s$^{-1}$ reported by \citet{Weilbacher03} and with same line width.
Additionally, we included  a one-degree polynomial to model the stellar continuum.
The DIB at $\lambda$5797.1 was not included in the fit since we  did not see any evidence for it  in our spectra.
It is known that the ratio of the equivalent width of these two DIBs can vary up to a factor $\sim$5, depending on the characteristics of the ISM in the sight line of consideration \citep[][]{Vos11}, but only  in very specific situations both DIBs are comparable. Typically,  the DIB $\lambda$5797.1 is  $\sim$3-5 times fainter.

 \begin{figure}[!th] \centering
 \includegraphics[width=0.43\textwidth, bb=65 60 525 520, clip=]{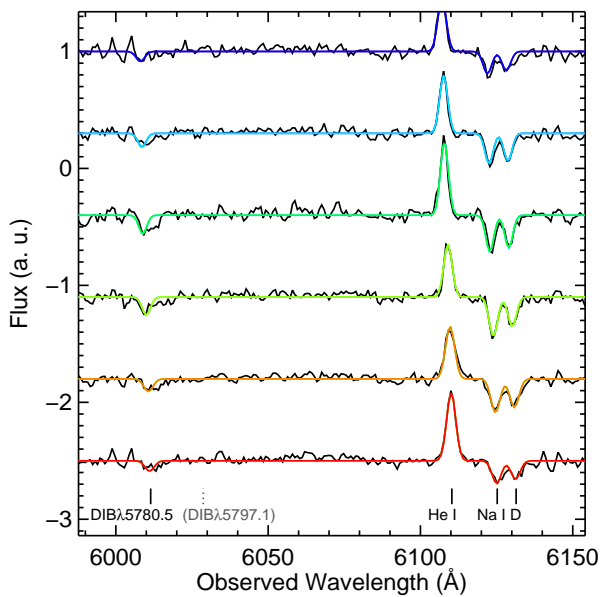} 
     \caption[Example of spectra]{Spectra for the six half rings showing the fit in the area of the DIB, ordered from smallest \emph{top} to largest \emph{bottom} velocity and normalized to the median in the displayed spectral range. 
     The expected locations of the discussed features  are marked at the bottom of the figure.
     The colour code in the fits echoes the ordering in velocity. Likewise, the shift of the spectral features is clearly visible.
        } 
    \label{figspec}
  \end{figure}

Figure \ref{figspec} shows the six fits for this critical range. The DIB at  $\lambda$5780.5 is clearly seen in all of them, with the possible exception of the outermost approaching half-ring. Hereafter, error bars indicate a preliminary estimate of the uncertainties as derived using Eq. 1 in \citet{MonrealIbero11}, which takes into account the signal-to-noise ratio of the spectra. However, the greatest source of uncertainty is the possible contamination in the spectra by a stellar  spectral feature at $\lambda$5782, rest frame,  with contributions from \ion{Fe}{i}, \ion{Cr}{i}, \ion{Cu}{i,} and \ion{Mg}{i} \citep{Worthey94}, and whose strength increases with age and metallicity.
An in-depth study to model this feature is beyond the scope of this initial experiment. In the following we discuss the expected maximum amount of contamination and provide an \emph{ad hoc} correction for it.

We measured  EW(\ha) in emission varying from $\sim90$~\AA\  in the inner rings to $\sim60$~\AA\  in the outer ones, which points to a very young (i.e. $\lsim$7~Myr) underlying stellar population.
The ages of the star-forming knots  in \object{AM\,1353-272} A \citep[\lsim40~Myr,][]{Weilbacher00}, as indicative of the moment when the interaction triggered the star formation would be another way to estimate the age of the stars in  \object{AM\,1353-272} B.
Likewise, using the $N2$ and $O3N2$ calibrators and the expressions provided by \citet{Marino13}, we estimate a metallicity for the inner rings of $12+\log(O/H)=8.50$ and 8.42, respectively (i.e. about half solar), decreasing outwards.
To estimate an upper limit for the contamination of the stellar feature, we conservatively created 1000 mock spectra by taking the spectrum of a 60~Myr single stellar population from the MILES library with solar metallicity \citep{FalconBarroso11}, adding Gaussians representing the DIBs at $\lambda$5780.5 and $\lambda$5797.1 and the  \hei$\lambda$5876 emission line. Then, we added noise with a scale of the standard deviation between 5\,020\,\AA\ and 5\,040\,\AA, rest-frame. EW($\lambda$5780.5) in the spectra was varied between 10~m\AA\, and 735~m\AA. As discussed above, a common range for the relative strength of the two DIBs is $\sim3-5$. Here, we assumed EW($\lambda$5797.1) = EW($\lambda$5780.5)/3. We measured EW($\lambda$5780.5) in these mock spectra using the same methodology as in the real data.  The relation between input and output equivalent widths was fitted to a one-degree polynomial that was used to estimate an upper limit to the correction due to  the stellar feature contamination. 
Measured EW($\lambda$5780.5)$\lsim200$\,m\AA\, are consistent with no DIB, EW($\lambda$5780.5)$\sim300$\,m\AA\, 
should be reduced by a 40\%. For EW($\lambda$5780.5)$\gsim500$\,m\AA\ a reduction of $\sim$20\% is enough. 
This simulation also suggests that under typical conditions, an EW($\lambda$5780.5)$\gsim$800~m\AA\, would be necessary to have some
expectations for detecting the DIB at $\lambda$5797.1.
 
\section{Results \label{secresults}}

\subsection{Radial profiles}

\begin{figure}[!htb] \centering
   \includegraphics[width=0.44\textwidth,  clip=]{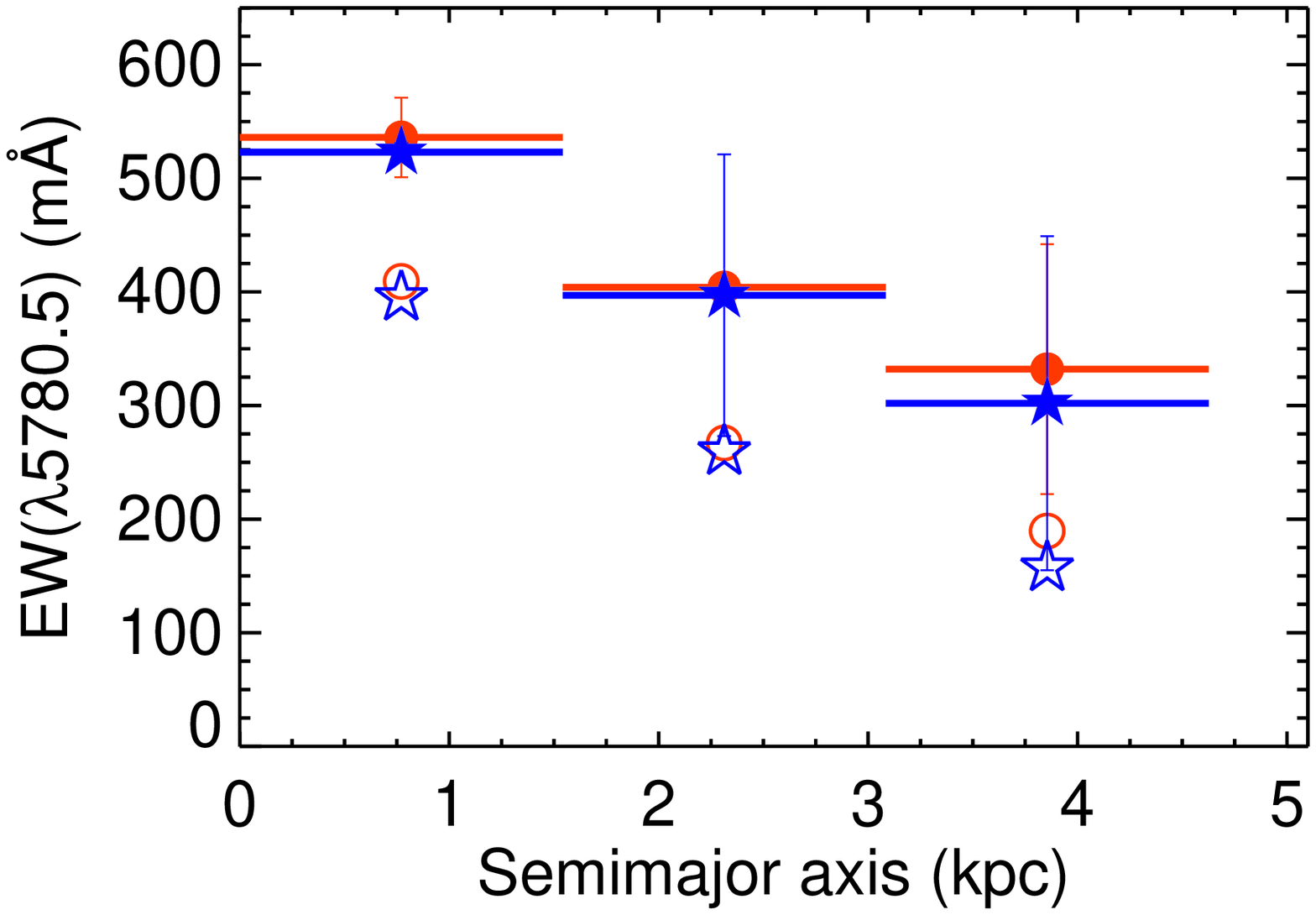} \\
   \includegraphics[width=0.44\textwidth,  clip=]{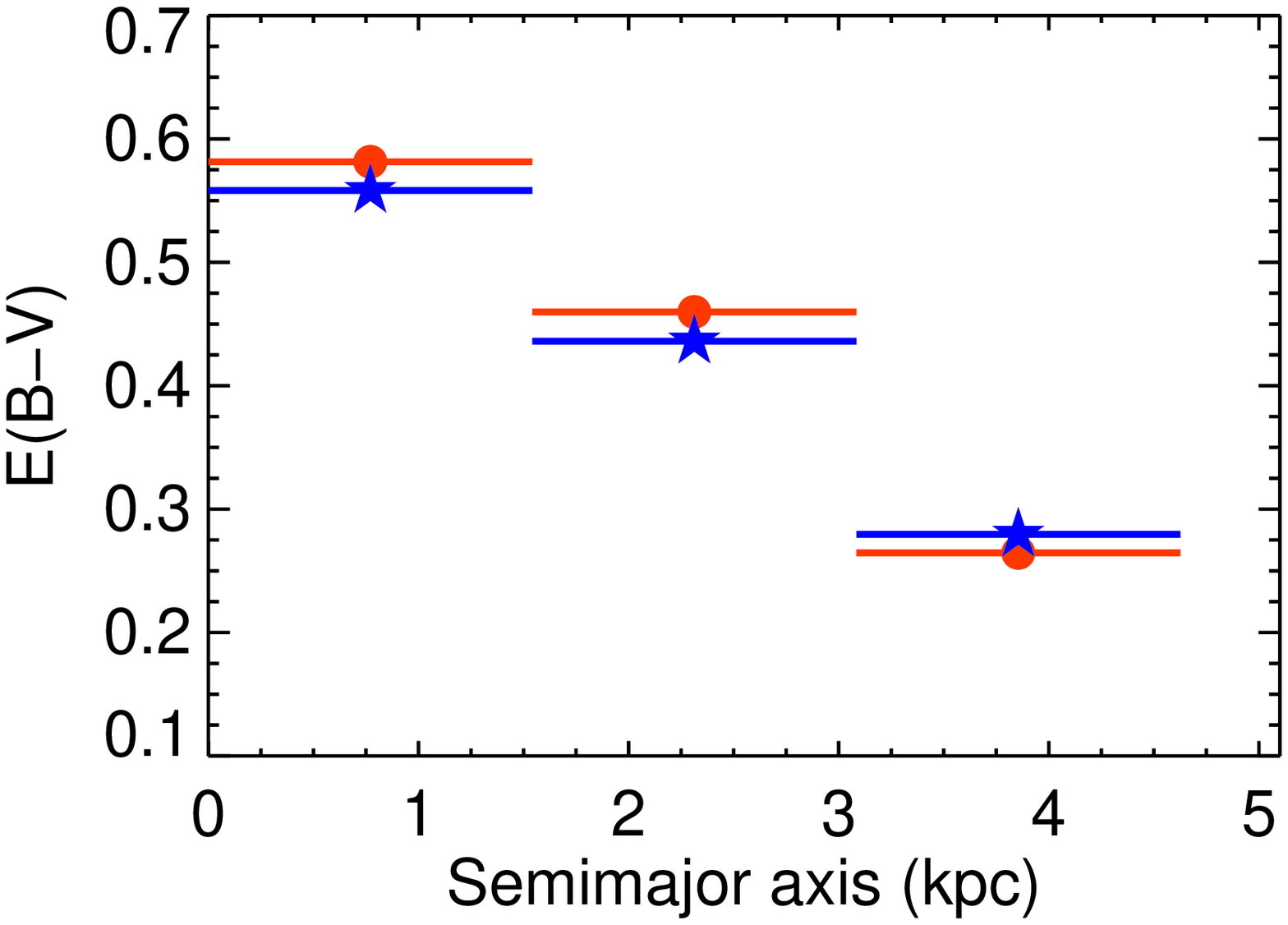} 
   \caption[Radial profiles]{Radial profiles for EW($\lambda$5780.5) (\emph{top}) and the extinction (\emph{bottom}) as derived from \ha/\hb\ for the approaching (\emph{stars and blue lines}) and receding  (\emph{circles and red lines}) sides. The horizontal bars traversing the filled symbols mark the radial extent over which the spectra were co-added. In the upper plot, filled symbols indicate measurements without correction, while open symbols include the maximum correction for the contamination of the stellar feature at $\lambda$5782\,\AA. 
}
      \label{figprofiles}
  \end{figure}

The upper plot of  Fig. \ref{figprofiles} shows  EW($\lambda$5780.5) as a function of the semimajor axis. The receding and approaching halves of the galaxy both present a negative gradient of the DIB strength with distance to the centre. This points towards a direct link with the intrinsic properties of the galaxy.  

One of the best established correlations between DIB strength and other characteristics of the ISM is that with the interstellar extinction \citep{Herbig93}. 
We derived an estimate of the extinction using the extinction curve of \citet{Fluks94} and assuming an intrinsic Balmer emission-line ratio \ha/\hb\ = 2.86 for a case B aproximation and $T_e = 10\,000$ K \citep{Osterbrock06}.
The corresponding radial profiles are shown in the lower plot of  Fig. \ref{figprofiles}. As  with the DIB, they display a decreasing gradient towards larger distances from the centre of the galaxy. This implies an intrinsic correlation between extinction and EW($\lambda$5780.5) for \object{AM1353-272}~ B and further supports the derived radial profiles for the DIB.

\subsection{Comparison with other sightlines}

\begin{figure}[!h] \centering
   \includegraphics[width=0.440\textwidth,  clip=]{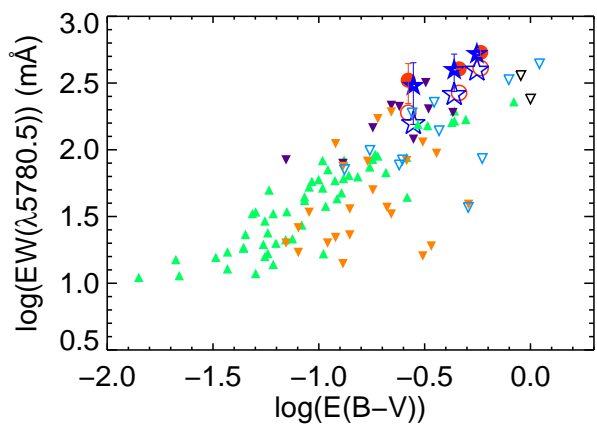} 
    \caption[EW(DIB) vs. E(B-V)]{Relation between EW($\lambda$5780.5) and extinction for \object{AM1353-272} (\emph{red circles} and \emph{blue stars}). 
Symbols and colours are as in Fig. \ref{figprofiles}. Additionally,
we show the following Galactic  (triangles)  and extragalactic (inverted triangles) sightlines:
     \emph{green} \citep[Milky Way,][]{Puspitarini13};
     \emph{purple} \citep[M31,][]{Cordiner11},
     \emph{orange} \citep[Magellanic Clouds,][]{Welty06},
    \emph{light blue} \citep[supernovae,][]{Phillips13},
    \emph{black} \citep[dusty starbursts,][]{Heckman00}.
    We used filled symbols when a galaxy had spatially resolved detections and open symbols when there is a single sightline for a given target.
}
    \label{figdibvsebv}
  \end{figure}

Figure \ref{figdibvsebv} illustrates the comparison between this intrinsic correlation with measurements  for other sightlines.
The sample is by no means complete, but covers a diversity of environments, methodologies, sampled areas, etc.  
This includes examples of spatially resolved measurements of the two large spirals in the Local Group \citep{Puspitarini13,Cordiner11} as well as for the Magellanic Clouds \citep{Welty06}. Likewise, we included some supernovae \citep{Phillips13} and nearby galaxies \citep{Heckman00}. 
There is a clear correlation between extinction and EW($\lambda$5780.5). Specifically, the ratio between this two quantities across \object{AM 1353-272} B is constant (within experimental errors), despite variations in metallicity, radiation field, etc. (Weilbacher et al. in prep).
When examining the ensemble of data globally, 
the dispersion is larger, pointing at secondary parameters modulating this correlation.
In particular, for a given reddening, DIBs in  the Magellanic Clouds are a factor 2-6 fainter than Galactic sightlines, as pointed out by  \citet{Welty06}.
All of our measurements in  \object{AM\,1353-272}\,B agree well with this general relation.
Interestingly enough, the relation between EW($\lambda$5780.5) and $E(B-V)$ is tighter when only resolved studies of disk-like galaxies are taken into account (Milky Way, Andromeda, our data). \object{AM\,1353-272}\,B indeed occupies an extension towards high EWs and $E(B-V)$ of the relation found for the large spirals in the Local Group.

\section{Summary and perpectives}

This work presents the first spatially resolved detection of a DIB in a galaxy outside the Local Group. By using 568 MUSE spectra sampling a total projected area of 13.5~kpc$^2$, we were able to measure the equivalent width of the DIB at $\lambda$5780.5 in six locations of \object{AM\,1353-272}\,B. Our strategy constitutes  an alternative approach to extragalactic DIB studies different from using targets in the Local Group and individual sightlines outside the Local Group. We found three main results:

\begin{enumerate}
      \item We found decreasing radial profiles for EW($\lambda$5780.5), both in the receding and approaching sides of \object{AM\,1353-272}\, B up to distances of $\sim$4.6~kpc from the galaxy centre.
      \item The interstellar extinction displays a similar trend, with decreasing values towards the external parts. This translates into a correlation between the strength of the DIB and the extinction within  \object{AM\,1353-272}\,B. 
      \item A comparison of $E(B-V)$ and EW($\lambda$5780.5) in \object{AM\,1353-272}\,B and other sightlines shows that this intrinsic correlation agrees with the existing global trend between these quantities, especially when compared with targets in the spiral galaxies of the Local Group (i.e. our Galaxy and M31).
\end{enumerate}

When seen globally, these three results demonstrate that the spatially resolved detection of DIBs in galaxies outside the Local Group is possible and 2D DIB mapping feasible  thanks to high-sensitivity instruments like MUSE. 
There are currently no 2D maps of DIBs in galaxies other than the Milky Way.
The construction of such maps will permit studying the relationship between the DIBs and other components of the ISM in conditions different from those found in our Galaxy. We will be able to
address questions like whether an extinction-DIB correlation in extremely dense environments also exists,
to which degree the strength of the different DIBs depends on the dust-to-gas ratio,
how DIBs react to the radiation field typical of an environment of extreme star formation, 
or 
whether the different relations locally found (e.g. the various DIB families) also exist in other galaxies, and how they depend on the intrinsic characteristics of the galaxies (e.g. morphology, metallicity, etc.).
 The  answers will be a critical test  that will help to constrain the  nature of the carriers.
  
\begin{acknowledgements}
We thank the referee for the diligent reading of the manuscript as well as for providing valuable comments that helped us to clarify and improve the first submitted version of this letter.
We thank the MUSE collaboration and PI R. Bacon for building this  extraordinary spectrograph and the rest of the commissioning  team, G. Zins in particular, for the tireless effort to continue  improving the instrument through all runs.
AM-I and RL acknowledge support from Agence Nationale de
la Recherche through the STILISM project (ANR-12-BS05-0016-02).
PMW and SK received support through BMBF Verbundforschung (project MUSE-AO,
grants 05A14BAC and 05A14MGA).
\\
\end{acknowledgements}

\bibliography{mybib_aa}{}

\begin{thebibliography}{35}
\expandafter\ifx\csname natexlab\endcsname\relax\def\natexlab#1{#1}\fi

\bibitem[{{Bacon} {et~al.}(2014){Bacon}, {Vernet}, {Borisiva}, {Bouch{\'e}},
  {Brinchmann}, {Carollo}, {Carton}, {Caruana}, {Cerda}, {Contini}, {Franx},
  {Girard}, {Guerou}, {Haddad}, {Hau}, {Herenz}, {Herrera}, {Husemann},
  {Husser}, {Jarno}, {Kamann}, {Krajnovic}, {Lilly}, {Mainieri}, {Martinsson},
  {Palsa}, {Patricio}, {P{\'e}contal}, {Pello}, {Piqueras}, {Richard},
  {Sandin}, {Schroetter}, {Selman}, {Shirazi}, {Smette}, {Soto}, {Streicher},
  {Urrutia}, {Weilbacher}, {Wisotzki}, \& {Zins}}]{Bacon14}
{Bacon}, R., {Vernet}, J., {Borisiva}, E., {et~al.} 2014, The Messenger, 157,
  13

\bibitem[{{Cordiner} {et~al.}(2011){Cordiner}, {Cox}, {Evans}, {Trundle},
  {Smith}, {Sarre}, \& {Gordon}}]{Cordiner11}
{Cordiner}, M.~A., {Cox}, N.~L.~J., {Evans}, C.~J., {et~al.} 2011, \apj, 726,
  39

\bibitem[{{Cordiner} {et~al.}(2013){Cordiner}, {Fossey}, {Smith}, \&
  {Sarre}}]{Cordiner13}
{Cordiner}, M.~A., {Fossey}, S.~J., {Smith}, A.~M., \& {Sarre}, P.~J. 2013,
  \apjl, 764, L10

\bibitem[{{Ehrenfreund} {et~al.}(2002){Ehrenfreund}, {Cami},
  {Jim{\'e}nez-Vicente}, {Foing}, {Kaper}, {van der Meer}, {Cox},
  {D'Hendecourt}, {Maier}, {Salama}, {Sarre}, {Snow}, \&
  {Sonnentrucker}}]{Ehrenfreund02}
{Ehrenfreund}, P., {Cami}, J., {Jim{\'e}nez-Vicente}, J., {et~al.} 2002, \apjl,
  576, L117

\bibitem[{{Falc{\'o}n-Barroso} {et~al.}(2011){Falc{\'o}n-Barroso},
  {S{\'a}nchez-Bl{\'a}zquez}, {Vazdekis}, {Ricciardelli}, {Cardiel}, {Cenarro},
  {Gorgas}, \& {Peletier}}]{FalconBarroso11}
{Falc{\'o}n-Barroso}, J., {S{\'a}nchez-Bl{\'a}zquez}, P., {Vazdekis}, A.,
  {et~al.} 2011, \aap, 532, A95

\bibitem[{{Fluks} {et~al.}(1994){Fluks}, {Plez}, {The}, {de Winter},
  {Westerlund}, \& {Steenman}}]{Fluks94}
{Fluks}, M.~A., {Plez}, B., {The}, P.~S., {et~al.} 1994, \aaps, 105, 311

\bibitem[{{Friedman} {et~al.}(2011){Friedman}, {York}, {McCall}, {Dahlstrom},
  {Sonnentrucker}, {Welty}, {Drosback}, {Hobbs}, {Rachford}, \&
  {Snow}}]{Friedman11}
{Friedman}, S.~D., {York}, D.~G., {McCall}, B.~J., {et~al.} 2011, \apj, 727, 33

\bibitem[{{Fulara} \& {Kre{\l}owski}(2000)}]{Fulara00}
{Fulara}, J. \& {Kre{\l}owski}, J. 2000, \nar, 44, 581

\bibitem[{{Heckman} \& {Lehnert}(2000)}]{Heckman00}
{Heckman}, T.~M. \& {Lehnert}, M.~D. 2000, \apj, 537, 690

\bibitem[{{Heger}(1922)}]{Heger22}
{Heger}, M.~L. 1922, Lick Observatory Bulletin, 10, 141

\bibitem[{{Herbig}(1993)}]{Herbig93}
{Herbig}, G.~H. 1993, \apj, 407, 142

\bibitem[{{Herbig}(1995)}]{Herbig95}
{Herbig}, G.~H. 1995, ARAA, 33, 19

\bibitem[{{Kos} {et~al.}(2014){Kos}, {Zwitter}, {Wyse}, {Bienaym{\'e}},
  {Binney}, {Bland-Hawthorn}, {Freeman}, {Gibson}, {Gilmore}, {Grebel},
  {Helmi}, {Kordopatis}, {Munari}, {Navarro}, {Parker}, {Reid}, {Seabroke},
  {Sharma}, {Siebert}, {Siviero}, {Steinmetz}, {Watson}, \& {Williams}}]{Kos14}
{Kos}, J., {Zwitter}, T., {Wyse}, R., {et~al.} 2014, Science, 345, 791

\bibitem[{{Marino} {et~al.}(2013){Marino}, {Rosales-Ortega}, {S{\'a}nchez},
  {Gil de Paz}, {V{\'{\i}}lchez}, {Miralles-Caballero}, {Kehrig},
  {P{\'e}rez-Montero}, {Stanishev}, {Iglesias-P{\'a}ramo}, {D{\'{\i}}az},
  {Castillo-Morales}, {Kennicutt}, {L{\'o}pez-S{\'a}nchez}, {Galbany},
  {Garc{\'{\i}}a-Benito}, {Mast}, {Mendez-Abreu}, {Monreal-Ibero}, {Husemann},
  {Walcher}, {Garc{\'{\i}}a-Lorenzo}, {Masegosa}, {Del Olmo Orozco},
  {Mour{\~a}o}, {Ziegler}, {Moll{\'a}}, {Papaderos},
  {S{\'a}nchez-Bl{\'a}zquez}, {Gonz{\'a}lez Delgado}, {Falc{\'o}n-Barroso},
  {Roth}, {van de Ven}, \& {Califa Team}}]{Marino13}
{Marino}, R.~A., {Rosales-Ortega}, F.~F., {S{\'a}nchez}, S.~F., {et~al.} 2013,
  \aap, 559, A114

\bibitem[{{Markwardt}(2009)}]{Markwardt09}
{Markwardt}, C.~B. 2009, in ASP Conf. Series, Vol. 411, ASP Conf. Series, ed.
  {D.~A.~Bohlender, D.~Durand, \& P.~Dowler}, 251--+

\bibitem[{{Merrill}(1934)}]{Merrill34}
{Merrill}, P.~W. 1934, \pasp, 46, 206

\bibitem[{{Monreal-Ibero} {et~al.}(2011){Monreal-Ibero}, {Rela{\~n}o},
  {Kehrig}, {P{\'e}rez-Montero}, {V{\'{\i}}lchez}, {Kelz}, {Roth}, \&
  {Streicher}}]{MonrealIbero11}
{Monreal-Ibero}, A., {Rela{\~n}o}, M., {Kehrig}, C., {et~al.} 2011, \mnras,
  413, 2242

\bibitem[{{Monreal-Ibero} {et~al.}(2005){Monreal-Ibero}, {Roth},
  {Sch{\"o}nberner}, {Steffen}, \& {B{\"o}hm}}]{MonrealIbero05}
{Monreal-Ibero}, A., {Roth}, M.~M., {Sch{\"o}nberner}, D., {Steffen}, M., \&
  {B{\"o}hm}, P. 2005, \apjl, 628, L139

\bibitem[{{Monreal-Ibero} {et~al.}(2010){Monreal-Ibero}, {V{\'{\i}}lchez},
  {Walsh}, \& {Mu{\~n}oz-Tu{\~n}{\'o}n}}]{MonrealIbero10a}
{Monreal-Ibero}, A., {V{\'{\i}}lchez}, J.~M., {Walsh}, J.~R., \&
  {Mu{\~n}oz-Tu{\~n}{\'o}n}, C. 2010, \aap, 517, A27+

\bibitem[{{Osterbrock} \& {Ferland}(2006)}]{Osterbrock06}
{Osterbrock}, D.~E. \& {Ferland}, G.~J. 2006, {Astrophysics of gaseous nebulae
  and active galactic nuclei}, ed. D.~E. {Osterbrock} \& G.~J. {Ferland}

\bibitem[{{Phillips} {et~al.}(2013){Phillips}, {Simon}, {Morrell}, {Burns},
  {Cox}, {Foley}, {Karakas}, {Patat}, {Sternberg}, {Williams}, {Gal-Yam},
  {Hsiao}, {Leonard}, {Persson}, {Stritzinger}, {Thompson}, {Campillay},
  {Contreras}, {Folatelli}, {Freedman}, {Hamuy}, {Roth}, {Shields}, {Suntzeff},
  {Chomiuk}, {Ivans}, {Madore}, {Penprase}, {Perley}, {Pignata}, {Preston}, \&
  {Soderberg}}]{Phillips13}
{Phillips}, M.~M., {Simon}, J.~D., {Morrell}, N., {et~al.} 2013, \apj, 779, 38

\bibitem[{{Puspitarini} {et~al.}(2015){Puspitarini}, {Lallement}, {Babusiaux},
  {Chen}, {Bonifacio}, {Sbordone}, {Caffau}, {Duffau}, {Hill}, {Monreal-Ibero},
  {Royer}, {Arenou}, {Peralta}, {Drew}, {Bonito}, {Lopez-Santiago}, {Alfaro},
  {Bensby}, {Bragaglia}, {Flaccomio}, {Lanzafame}, {Pancino}, {Recio-Blanco},
  {Smiljanic}, {Costado}, {Lardo}, {de Laverny}, \& {Zwitter}}]{Puspitarini15}
{Puspitarini}, L., {Lallement}, R., {Babusiaux}, C., {et~al.} 2015, \aap, 573,
  A35

\bibitem[{{Puspitarini} {et~al.}(2013){Puspitarini}, {Lallement}, \&
  {Chen}}]{Puspitarini13}
{Puspitarini}, L., {Lallement}, R., \& {Chen}, H.-C. 2013, \aap, 555, A25

\bibitem[{{Rieke} \& {Lebofsky}(1985)}]{Rieke85}
{Rieke}, G.~H. \& {Lebofsky}, M.~J. 1985, \apj, 288, 618

\bibitem[{{Sandin} {et~al.}(2008){Sandin}, {Sch{\"o}nberner}, {Roth},
  {Steffen}, {B{\"o}hm}, \& {Monreal-Ibero}}]{Sandin08}
{Sandin}, C., {Sch{\"o}nberner}, D., {Roth}, M.~M., {et~al.} 2008, \aap, 486,
  545

\bibitem[{{Schlafly} \& {Finkbeiner}(2011)}]{Schlafly11}
{Schlafly}, E.~F. \& {Finkbeiner}, D.~P. 2011, \apj, 737, 103

\bibitem[{{van Loon} {et~al.}(2013){van Loon}, {Bailey}, {Tatton}, {Ma{\'{\i}}z
  Apell{\'a}niz}, {Crowther}, {de Koter}, {Evans}, {H{\'e}nault-Brunet},
  {Howarth}, {Richter}, {Sana}, {Sim{\'o}n-D{\'{\i}}az}, {Taylor}, \&
  {Walborn}}]{vanLoon13}
{van Loon}, J.~T., {Bailey}, M., {Tatton}, B.~L., {et~al.} 2013, \aap, 550,
  A108

\bibitem[{{Vos} {et~al.}(2011){Vos}, {Cox}, {Kaper}, {Spaans}, \&
  {Ehrenfreund}}]{Vos11}
{Vos}, D.~A.~I., {Cox}, N.~L.~J., {Kaper}, L., {Spaans}, M., \& {Ehrenfreund},
  P. 2011, \aap, 533, A129

\bibitem[{{Weilbacher} {et~al.}(2003){Weilbacher}, {Duc}, \&
  {Fritze-v.~Alvensleben}}]{Weilbacher03}
{Weilbacher}, P.~M., {Duc}, P.-A., \& {Fritze-v.~Alvensleben}, U. 2003, \aap,
  397, 545

\bibitem[{{Weilbacher} {et~al.}(2000){Weilbacher}, {Duc}, {Fritze
  v.~Alvensleben}, {Martin}, \& {Fricke}}]{Weilbacher00}
{Weilbacher}, P.~M., {Duc}, P.-A., {Fritze v.~Alvensleben}, U., {Martin}, P.,
  \& {Fricke}, K.~J. 2000, \aap, 358, 819

\bibitem[{{Weilbacher} {et~al.}(2002){Weilbacher}, {Fritze-v.~Alvensleben},
  {Duc}, \& {Fricke}}]{Weilbacher02}
{Weilbacher}, P.~M., {Fritze-v.~Alvensleben}, U., {Duc}, P.-A., \& {Fricke},
  K.~J. 2002, \apjl, 579, L79

\bibitem[{{Weilbacher} {et~al.}(2014){Weilbacher}, {Streicher}, {Urrutia},
  {P{\'e}contal-Rousset}, {Jarno}, \& {Bacon}}]{Weilbacher14}
{Weilbacher}, P.~M., {Streicher}, O., {Urrutia}, T., {et~al.} 2014, in
  Astronomical Society of the Pacific Conference Series, Vol. 485, Astronomical
  Data Analysis Software and Systems XXIII, ed. N.~{Manset} \& P.~{Forshay},
  451

\bibitem[{{Welty} {et~al.}(2006){Welty}, {Federman}, {Gredel}, {Thorburn}, \&
  {Lambert}}]{Welty06}
{Welty}, D.~E., {Federman}, S.~R., {Gredel}, R., {Thorburn}, J.~A., \&
  {Lambert}, D.~L. 2006, \apjs, 165, 138

\bibitem[{{Worthey} {et~al.}(1994){Worthey}, {Faber}, {Gonzalez}, \&
  {Burstein}}]{Worthey94}
{Worthey}, G., {Faber}, S.~M., {Gonzalez}, J.~J., \& {Burstein}, D. 1994,
  \apjs, 94, 687

\bibitem[{{Zasowski} {et~al.}(2015){Zasowski}, {M{\'e}nard}, {Bizyaev},
  {Garc{\'{\i}}a-Hern{\'a}ndez}, {Garc{\'{\i}}a P{\'e}rez}, {Hayden},
  {Holtzman}, {Johnson}, {Kinemuchi}, {Majewski}, {Nidever}, {Shetrone}, \&
  {Wilson}}]{Zasowski15}
{Zasowski}, G., {M{\'e}nard}, B., {Bizyaev}, D., {et~al.} 2015, \apj, 798, 35

\end{thebibliography}
\bibliographystyle{./aa}

\end{document}